\documentclass[12pt]{article}
 
\usepackage{amssymb} 
  \def\({\left(}       \def\){\right)}    \def\6{\partial }  
  \def\lk{\,\left[ \,} \def\rk{\,\right] \,} 
         \def\0{\over }    
\def\ov#1{\overline{#1}}   
\def\P{ {\mit\Pi} } \def\ln{{\rm ln}} \def\Tr{{\rm Tr}}  
 \let\a=\alpha \let\b=\beta \let\d=\delta \let\e=\varepsilon
 \let\l=\lambda \let\o=\omega \let\s=\sigma \let\D=\Delta  
\def\eq#1{(\ref{#1})}  \def\nonu{\nonumber}
\def\be#1{\begin{equation} \label{#1}} \def\ee{\end{equation}} 
\def\bea#1{\begin{eqnarray} \label{#1}} \def\eea{\end{eqnarray}}
\let\thq=\theequation   \def\ft{\footnotesize}
  \def\vc#1{{\buildrel _\rightharpoonup \over #1}}
  \def\pfeil{_\rightharpoonup} 
  \def\vcsm#1{ \def\sm{\raise 1.6pt\hbox to 5pt{\hss $_#1$}} 
               {\buildrel \pfeil \over \sm}}
\font\sf=cmss12 \def\MB{\hbox{\sf B}} \def\MD{\hbox{\sf D}}    
\font\log=logo10 scaled \magstep2 \def\MA{\hbox{\log A}}
\def\lamm{\,[\, p^2 - {(\vcsm q \vcsm p)^2 \0 q^2} ]\,}
\def\lammi{\,[\, p^2 - {(\vcsm q \vcsm p)^2 \0 ^{q^2}} ]\,}
\def\Dm{{\D^{\hspace{-.04cm}-}\hspace*{-.14cm}}}

\begin{document}


\begin{titlepage}  \begin{flushleft}  
ITP--UH $\;15 \, / \, 99$ \\
hep-ph/9909551 \end{flushleft} \vspace{-1.2cm}
   \begin{flushright} September 1999 \hspace{.7cm} $ $ \\
   \end{flushright} \vfill \vskip 2cm 
   \begin{center} {\Large \bf 
Magnetic screening in the hot gluon system} \\  
\vfil  \vspace{2cm} 
\renewcommand{\thefootnote}{\fnsymbol{footnote}} 
  {\large 
Jens Reinbach \ and \ Hermann Schulz
   \footnote[1]{~E--mail~: ~reinbach@itp.uni-hannover.de~, 
                           ~hschulz@itp.uni--hannover.de } }
   \renewcommand{\thefootnote}{\arabic{footnote}} 
   \\[6pt] \bigskip  {\sl    
Institut f\"ur Theoretische Physik,
Universit\"at Hannover \\
Appelstra\ss e 2, D-30167 Hannover, Germany \\  }
    \vspace{2cm} \vfill  {\bf    
Abstract \quad }  \end{center} \begin{quotation}  \ \  
  The gluon transverse self--energy of the pure 
  Yang--Mills system at high--temperature
  is analysed in the static limit and at fourth 
  order in the coupling. Possible contributions to 
  this function are collected, seen to be gauge--fixing 
  independent subsets and shown to vanish all, except 
  those which are either regulators or constituents of
  the self--energy of Euklidean 3D Yang--Mills theory at 
  zero temperature. The latter self--energy, in turn, is 
  known from the non--perturbative analysis by Karabali 
  and Nair.

\vspace{.8cm} \noindent PACS numbers : \ 
 11.10.Wx , \ 11.15.Bt  
\vspace*{.5cm}
\end{quotation}
\end{titlepage}

$ $ 

{\bf 1. Introduction}

Twenty years ago it was observed by Linde \cite{linde} that the 
perturbative treatment of the high--temperature Yang--Mills 
system runs into a serious problem. If a magnetic mass $\tau$, 
the system might be able to generate thermally, falls short of 
$g^2T$ in magnitude, the series would diverge, and a phase of 
deconfined gluons could not exist. But even if $\tau \sim g^2T$,
the perturbation series becomes an (unknown) numerical series. 
Due to this phenomenon \cite{gpy}, no one was able so far to 
calculate the pressure at order $g^6$ or the gluon self-energy 
at $g^4$ \ -- \ a shame for analytical theoretical physics. Today, 
however, there is a way out. It is provided by the non-perturbative 
analysis of Karabali, Kim and Nair (referred to as KKN in the sequel) 
of 2+1D Yang--Mills theory at zero temperature \cite{knvor,kn,knnach}. 
In this note, by studying static magnetic screening, which is 
presumably the simplest example, it is shown how to use KKN's 
results in an otherwise perturbative treatment.   

The ''Linde sea'' of diagrams is easily understood from figure 1. 
If one more line is added to an arbitrary skeleton diagram, e.g.
in the manner shown in the figure, then, in the sense of power 
counting, it has two more 3--vertices ($\sim p^2 g^2$), three 
more propagators ($\sim (p^2 + \tau^2)^{-3}$) and one more loop 
integration ($T\int\! d^3p$, if reduced to the term with zero 
Matsubara frequency). Thus, the (n+1)--loop and n--loop 
differ by a factor $\sim g^2 T \int\! d^3p \, p^2 (p^2+\tau^2)^{-3} 
\sim g^2T/\tau\,$. For $\tau \sim g^2 T$ this factor has order 1 
in magnitude. Once the zero--frequency modes become relevant, 
{\it all} skeletons contribute with the same order of magnitude. 
Any finite--n--loop calculation of the magnetic mass is thus 
inconsistent. 

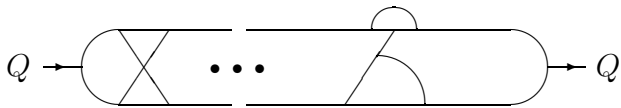
\begin{figure}[bth]  \unitlength 1cm  
\hspace*{2cm} \begin{picture}(11,1)
 \put(4.5,.3){\oval(4,1)[l]}  \put(4.7,.3){\oval(8,1)[r]}   
 \put(2,.3){\line(1,0){.5}}  \put(8.7,.3){\line(1,0){.5}}  
 \put(4.2,.28){$_\bullet$}     \put(4.5,.28){$_\bullet$}
 \put(4.8,.28){$_\bullet$}
\put(3,.8){\line(2,-3){.66}} \put(3,-.2){\line(2,3){.66}}
\put(6.66,.8){\oval(.6,.6)[t]} \put(6,-.2){\line(2,3){.66}}
\put(6.42,-.2){\oval(1.3,1.3)[tr]} 
\put(1.5,.19){$Q$}  \put(2,.3){\vector(1,0){.3}} 
\put(9.34,.19){$Q$} \put(8.8,.3){\vector(1,0){.3}}
\end{picture}
\caption{\ft An arbitrary 2--leg n--loop skeleton diagram
   with one line added$\,$: the half circle on top, say, or,
   equivalently, the one below. The outer momentum $Q$ 
   is static ($Q_0=0$) and supersoft ($q \sim\tau$).}
\end{figure}

Bosonic fields live on a cylinder with circumference $\b=1/T$. 
Each loop integration $\sum_P \equiv T\sum_n \int^3_p$, $\int^3_p 
\equiv (2\pi)^{-3} \int\! d^3p$, has its zero--frequency part 
$T\int^3_p\,$. A field depen\-ding on $P$ looses dependence on its 
time coordinate in this part. Irrespective of the physical 
quantity under study, the subset of contributions with $P_0=0$ in 
{\sl all} loops might be the full set of an Euklidean physics at 
$T=0$ in three dimensions \cite{appel,brani,nie}. However, 
this theory needs regulators to be derived from the underlying 
4D setup.  

All about the {\it regularized} 3D Euklidean Yang--Mills theory is 
inherent in its 2+1D version, and this is the system treated by 
KKN. For an {\sl outline of the main argument} see \cite{kn}.
Appropriate shortcuts are found in \cite{knproc}, and for a 
low--level introduction see \cite{lecture}. Working in Weyl gauge and 
after splitting off the remaining gauge volume, KKN quantize the 
physical degrees of freedom in the Schroedinger functional picture.
A point--splitting regularization enables them to write scalar 
products of functional states as correlators of the hermitean WZW 
model. Conformal field theory then reveals the structure of 
normalizable wave functionals. Eigenstates of the Hamiltonian 
are constructed. The mass in the 3D gluon propagator is found 
to have the value $\,g^2 N T /(2\pi)\,$ at its leading order. 

While taking KKN's results for granted, the subject of this letter 
is the perturbation theory for magnetic screening, i.e. for the 
static limit of a ''dynamical'' quantity. Dynamics rests on 
linear response theory. The response of the gluon medium to 
infinitesimal perturbations is obtained by analytical continuation 
$Q_0 \to \o + i\e$ from Matsubara frequencies $Q_0=i 2\pi n T\,$. 
The breakthrough in understanding the gluon dynamics came in 1990, 
because the ''zeroth approximation'' of high--$T$ QCD was 
established only then \cite{bp,fret} and given the form of an 
effective action \cite{tawo,freta,bpeff,efran}. In the whole 
$\o$--$\vc q$ space gauge fixing independence is garantied 
\cite{kokure} only along the dispersion lines, longitudinal and 
transverse. In a diagram $\o^2$ over $q^2$, which is 
\hbox{figure 1} in \cite{frosch}, the transverse line 
runs down near the light--cone but then deviates from the 
Dirac--spectrum shape to reach the plasmon frequency 
$m^2 = g^2T^2/9$ at $q^2=0$. Crossing this point, $q^2$ becomes 
negative. The transverse line then changes slope and turns towards 
the origin. 

The static limit refers to zero frequency, i.e. to the lower 
border of the $\o^2>0$ half--plane. According to the above, only 
two points on this border line have physical meaning, namely the 
end points $-q^2=\s^2$ (Debye screening) and $-q^2=\tau^2$ 
(magnetic screening) of the two dispersion lines. Here, $\s^2$ 
is the squared Debye mass including its correction $\sim g^3$ 
\cite{reb}. Its leading term is $3m^2$. The 
transverse polarization function $\P_t(Q)$, on the other hand, 
vanishes at $Q_0=0$ to leading order ($\sim g^2$) as well as in 
next--to--leading order \cite{reb,frosch}, thereby giving rise to 
the magnetic mass problem. It is resolved at order $g^4$ where 
the solution $-q^2=\tau^2$ to the equation $0=q^2+\P_t(Q_0=0, q)$ 
attains a positive value. 

The purpose of this note is threefold: \\[2pt]
1. ~\parbox[t]{14.6cm}{Though it is rather plausible, 
    that the 4D object $\P_t(Q_0=0,q)$ may be identified with 
    $\P_t(q)$ of the zero--temperature 3D Yang--Mills system, 
    we like to remove possible doubts in this step.} \\[8pt]
2. ~\parbox[t]{14.6cm}{It has to be seen with detail how diagrammatic 
   contributions reduce to the correct 3D  Euklidean rest.} \\[8pt]
3. ~\parbox[t]{14.6cm}{Regulators are to be prepared out of 
    the 4D thermal theory.}   

We shall analyse $\P_t(Q_0=0,q)$ up to order $g^4$ only. Momenta of 
order $T$, $gT$, $g^2T$ in magnitude are called hard, soft and 
supersoft, respectively. Then, with an upper--left index denoting 
the number of loops,
\be{1}
   - q^2 = \; ^1\hspace*{-.8mm}\P_t^{\rm bare} 
          +\, ^1\hspace*{-.8mm}\P_t^{\rm next-to} 
          +\, ^2\hspace*{-.8mm}\P_t^{\rm hard-hard}
          +\, ^2\hspace*{-.8mm}\P_t^{\rm hard-supersoft} 
          +\; ^{\geqslant 2}\hspace*{-.7mm}\P_t^{\,
  \raise 1pt\hbox{\scriptsize all momenta supersoft}} \quad 
\ee 
is all we have to include. Diagrams with more than two loops enter 
through the Linde mechanism, i.e. only if all momenta are supersoft. 
Hence, the above last term is the two-- and higher loop part of the 
3D Euklidean theory. Its 1--loop pieces are to be taken from the 
first two terms of \eq{1}. But once they are combined with the last 
term, nothing might be left in the spirit of the above conjecture. 
In fact, after such regroupings, our final result may be written as 
\be{2}
  - q^2 \; = \;\; 0_{g^2}\;\; +\; 0_{g^3}\;\; +\; 0_{g^4}\;\;  
  + \; \P_{t\, ,\;{\rm 3D},\;T=0}^{\,
  \raise 2pt\hbox{\scriptsize regularized}} \;\; . \quad
\ee 
where each number zero corresponds to a gauge--fixing independent 
subset, and the index refers to the $g$ power to which it vanishes. 
As the 3D theory is a physics by itself, even the non--vanishing 
term of \eq{2} refers to a gauge--fixing independent subset. 

In sections 2, 3, 4 and 5 the first four terms of \eq{1} are 
studied, respectively.

$ $   

{\bf 2. One loop diagrams with bare lines}

While re--examining the transverse polarization function
$\P_t$ at one--loop order we shall take care of its $g^4$ 
contributions.

Our metrics is $+--\,-\;$. We shall need three members 
of the four--fold Lorentz matrix basis \cite{gpy}: 
\be{3}
 \MA= {(0,\vc p)\circ(0,\vc p)\0p^2} + g - U\circ U 
  \;\; , \;\; \MB= {(p^2,P_0 \vc p) \circ 
   (p^2,P_0 \vc p)\0 - P^2p^2 } 
  \;\; , \;\; \MD={P\circ P\0P^2} \;\; , \quad
\ee 
where $U=(1,\vc 0)$. The Lagrangian is written as
\be{4}
 {\cal L} =  - {1 \over  4} \,
 F_{\mu\nu}^{\,\;\;\; a} F^{\mu\nu \; a}
 - {1 \over  2 \alpha} \,\( \6^\mu A_\mu^a \)^2 
 + {1\02} A^{\mu\,a} (Y A)^a_\mu 
 + \, c_1^a \6^\mu  D_\mu^{ab} c_2^b 
 -  {1\02} A^{\mu\,a} (\ov{Y} A)^a_\mu  
\quad . \;\;
\ee 
with $F_{\mu\nu}^{\,\;\;\; a} = \6_\mu A_\nu^a
- \6_\nu A_\mu^a + g f^{abc} A_\mu^b A_\nu^c\,$ 
and $D_\mu^{ab} = \d^{ab} \6_\mu - g f^{abc} A_\mu^c\,$.
Renormalization is understood to be done at $T=0$. For the 
present, we may account for $Z$ factors, which are missing 
in \eq{4}, by omitting non--Bose--function parts in the
evaluation of frequency sums. 

At the right end of \eq{4} the mass term $\ov{Y}$ is 
subtracted to reinstall the original theory. The bare only 
indicates its use at one loop higher order. Through 
$(YA)^{\mu\,a} = \sum_P e^{-iPx}$ $Y^{\mu\nu}(P) A^a_\nu(P)$ 
this mass may be chosen momentum dependent \cite{rs}. By 
\be{5}
  Y^{\mu\nu}(P) \, = \, \MA^{\mu\nu}(P)\;\tau^2 \,\d_{P_0,0} 
  \; + \; \MB^{\mu\nu}(P)\;\s^2 \,\d_{P_0,0} \qquad
\ee 
we adopt the minimal version of infrared regularisation, 
which was found useful parti\-cularly in thermodynamics 
\cite{azi,pasi}. We are allowed to work with simple mass terms, 
in place of the full effective action \cite{bpeff}, 
because there is no hard--thermal--loop dressing of vertices at 
$\o=0$. The Kronecker version \eq{5} of these masses 
needs restriction to the static limit as well. 
In principle, in place of $\s^2$ there is some function of 
$p$ to be determined consistently. But we know from \cite{rs} 
that it becomes the longitudinal self--energy to the order of 
interest. Even $\tau^2$ is
inserted into \eq{5} as the expected outcome by summing
over the Linde sea. The possible dependence of $\tau^2$ 
on $p$ is of no relevance in the sequel. But according to 
KKN, and anticipating the announced three zeros, there 
might be no such dependence, at least not at supersoft $p$. 

The one--loop diagrams for $\P^{\mu\nu}$ are tadpole, loop and
ghost--loop. With view to \eq{5} it would be natural to
decompose $\sum_{P_0}$ into its $P_0=0$ and $P_0\neq 0$ part. 
However, two divergent pieces would arise this way. Therefore,
but also to keep contact with earlier work 
\cite{nt,flesh,frosch,flech}, we better distinguish bare from 
dressed lines, start with bare ones and defer the difference 
(giving $g^3$ terms) to the next section.

With bare gluon lines $G_0(P) = g^{\mu\nu}/P^2 + (\a-1)P^\mu 
P^\nu/P^4$, the transverse function $\P_t={1\02}\Tr 
\(\MA \P\)$, is conveniently split into three terms
\be{6}
  ^1\hspace*{-.8mm}\P_t^{\rm bare} =  
   \P_{(0)}  + \P_{(1)} + \P_{(2)} 
  \quad .
\ee 
For the last term see \eq{12} below.
The familar loop integration
\be{7}
  \P_{(0)} + \P_{(1)}   = g^2 N \sum_P \( - {2\0 P^2} 
  - {\, 2 \lamm \0 P^2 (P-Q)^2 } \)  \qquad
\ee 
shares the ''true zeroth order'' of the hot gluon system: 
$\P_{(0)}\sim g^2$ sets the scale. Watching $g^4$
terms, however, \eq{7} is not identical with its leading
order part. The latter,
\be{8}
   \P_{(0)} = {g^2 N T^2 \0 6}\;
   \lk  z^2 - (z^2-1) {z\02} \,\ln \( {z+1 \0 z-1} \) \rk
   \quad , \quad z \equiv {Q_0 \0 q} \quad , \quad
\ee 
is the well known transverse self--energy at leading order 
(see e.g.~\cite{nt}, Appendix B). It is gauge--fixing 
independent, and for $Q_0 \to \o = 0$ ($z=0$) it 
clearly vanishes. Let this zero contribution be 
the first number zero in equation \eq{2}.

As announced, \eq{8} is not \eq{7}. With a bit analysis, 
and turning to the static limit, we obtain
\be{9}
  \P_{(1)}^{Q_0=0} =  g^2 N {T \, q \0 16}
  \; = \; g^2 N T \int^3_p \( {4\03}\, {1\0p^2} - 
  {\, 2 \lamm \0 p^2 (\vc p - \vc q)^2 } \) \quad . \quad
\ee 

For the right 3D Euklidean Yang--Mills theory we must 
beware the $P_0=0$ part of \eq{7}. But if taken
at $Q_0=P_0=0$, $\sum \to T \int^3_p\,$, \eq{7} diverges. 
The integral in \eq{9}, on the other hand, is convergent, 
but the first term comes with the wrong factor. The 
resolution to this puzzle is by adding to \eq{9}
the following identity, valid for all $M\;$:
\be{10}  
  0 = \int^3_p \, {\,2 M^2\0 3\, p^2} \; \6_p \, 
      {p\0 M^2+p^2}  \; = \; \int^3_p \( 
       {2\03}\,{1\0p^2} - {2\0 M^2 + p^2} + 
     {4\03} {p^2 \0 (M^2+p^2)^2 } \) \quad . \quad
\ee 
Then \eq{9} turns into
\be{11}
  \P_{(1)}^{Q_0=0} =  g^2 N T \int^3_p 
     \( {2\0p^2}  - {2\0 M^2 + p^2} 
    - {\, 2 \lamm \0 p^2 (\vc p - \vc q)^2 }  
    + {4\03} {p^2 \0 (M^2+p^2)^2 } \) \;\; . \;
\ee 
Note that, for large $p$ and due to  angular 
integration, the square bracket may be replaced by 
$2p^2/3$. Hence, the terms to be preserved are supplied
with regulation in the UV. Obviously, apart from 
$M\gg g^2 T$, the value of the regulator mass $M$ may be 
chosen at will. 

There remains to notice the third contribution
to $\P_t^{\rm bare}\,$:
\bea{12}
 \P_{(2)} \!&=&\! g^2 N \sum_P \Bigg( \,
   {2\, Q^2 \0 P^2 (P-Q)^2 }  
   + \, (\a-1) \bigg[
       - {Q^2 \0 P^4 } +  { Q^4 \0 P^4 (P-Q)^2 } 
       - {\lamm Q^2 \0 P^4 (P-Q)^2 } \bigg] \nonu \\
  & & \hspace*{4.1cm} 
       - \,\; (\a-1)^2 \;{\,\lamm Q^4 \, \0 4 \, P^4 (P-Q)^4 }
         \;\Bigg) \quad . \quad
\eea 
Such terms are known from the next--to--leading
order analysis at soft scale. There, for the order $g^3$, 
they are all irrelevant \cite{nt} (\S 4 there). But here, 
where terms of order $g^4T^2$ are to be maintained, they 
are all relevant \ --- \ in spite of the supersoft 
$Q^2=-q^2$ in the numerators. This fact is easily realized 
by power counting with a supersoft IR regulator $\sim \tau$
in mind. Nonzero $P_0$ make $\P_{(2)} \sim g^6\,$. So, the
Matsubara sum in \eq{12} may be reduced to its $P_0=0$ term, 
hence $\sum$ replaced by $T\int^3_p$, immediately.

$ $   

{\bf 3. One loop~: the next--to--leading order}

Contributions of order $g^3$ to $\P_t$ (its ''true first 
approximation'') arise from 1-loop diagrams due dressing 
of lines and vertices \cite{bp,nt,flesh,frosch}. 
They are prepared by forming the difference
\be{13}
 ^1\hspace*{-.8mm}\P_t^{\rm next-to} 
  = \, ^1\hspace*{-.8mm}\P_t^{\rm dressed}   
  - \, ^1\hspace*{-.8mm}\P_t^{\rm bare} \quad . \quad   
\ee 
Thanks to the static limit (no vertex dressing) and to 
the economical IR regularization \eq{5}, terms with 
non--zero $P_0$ drop out in this difference. 
Moreover, the outer momentum has $Q_0=0$. Hence, 
\be{14}
  G^{\mu\nu}(P_0=0, \vc p) \; = \; - \MA^{\mu\nu}_{P_0=0} \,\D_\tau 
      - \MB^{\mu\nu}_{P_0=0} \,\D_\s 
      - \a\, \MD^{\mu\nu}_{P_0=0} \,\D_0    
\ee 
is all we need about the 4D dressed propagator. In \eq{14}, 
the 4D partial propagators (such as e.g.~$1/( P^2 - \s^2 
\d_{P_0,0}$) have turned \ -- \ under change of sign \ -- \
into the 3D Euklidean propagators 
\be{15}
  \D_\s = {1\0 p^2 + \s^2}  \;\, , \;\,
  \D_\tau = {1\0 p^2 + \tau^2}  \;\, , \;\,
  \D_0 = {1\0 p^2 } \;\; \hbox{and} \;\;
  \Dm_\s = {1\0 (\vc p - \vc q)^2 + \s^2} 
   \;\; \hbox{etc.} \;\; . \;
\ee 
The calculation of $\,^1\hspace*{-.8mm}\P_t^{\rm next-to}$
is straightforward, although a bit lengthy. It is much
simplified by the following four observations: \\
(a) Since not dressed, the ghost--loop needs not be
    included. \\
(b) At $P_0=0$ the matrix $\MB(P)$ becomes 
    $U \circ U$. Hence, as the matrix $\MA$ has 
    no zeroth components (moreover: $\MA_{P_0=0}=\MA$), 
    products of $\MA$ with $\MB$ vanish. \\
(c) Terms $\sim Q^\mu$, $\sim Q^\nu$ or $\sim B^{\mu\nu}$ may be 
    omitted, since $\P^{\mu\nu}$ will be traced with $\MA(Q)$, which
    is a projector with respect to $Q$. \\
(d) For inner momenta we have  $\MA(P) P = 0$ and 
    $\MA(P-Q) P = \MA(P-Q) Q\,$.  

According to (b) and (c) only very few terms with
$\D_\s$ survive, namely one in the tadpole diagram
(due $B^\l_\l = 1$) and one in the loop (due $U^\l 
U_\rho U_\l U^\rho =1$). The \hbox{result is}
\bea{16}
 \,^1\hspace*{-.8mm}\P_t^{\rm next-to}\!\! 
 &=&\! g^2 N T \int^3_p \, \Bigg\{ \;\D_\s - \D_0 
     - \lammi \( \Dm_\s \D_\s - \Dm_0 \D_0 \) 
     +  {4\03} \( \D_\tau - \D_0 \) \nonu \\
 & & \hspace*{-.1cm} 
      - \; {1\04} \lammi \bigg( \; 6   
      +  2 p^2 \Dm_0 + 12 q^2 \D_0 + q^4 \Dm_0 \D_0 
     \bigg) \( \Dm_\tau \D_\tau - \Dm_0 \D_0 \, \)  
     \nonu \\
 & &  \hspace*{-.1cm}
    + \;\a \, \Big( 2\lammi \D_0 -1 \Big)
      (p^2-q^2)^2 \,\Dm_0 \D_0 \( \D_\tau - \D_0 \)
      \;\;\Bigg\} \quad . \quad
\eea 

As always, we first look, whether the subset \eq{16} 
depends on the gauge--fixing parameter. It does. 
However, thanks to the last factor, the term with $\a$ 
is of order $g^2 T \int^3_p ( \D_\tau - \D_0 ) \sim 
g^2T\tau \sim g^4T^2$ in magnitude. Hence, the $g^3$ 
contribution to $\P_t$ at $Q_0=0$, which is fully 
contained in \eq{16}, is gauge--fixing independent. 

The $g^3$ subset even vanishes. To rederive this known fact 
\cite{reb} from \eq{16}, we select its $g^3$ terms as
\be{17}
 \,^1\hspace*{-.8mm}\P_t^{\rm next-to}\,\bigg|_{g^3} 
 =  g^2 N T \int^3_p \, \( 
   \D_\s - \D_0 - {2\03} p^2 \( \D_\s \D_\s - \D_0 \D_0 \) \)
 = 0 \quad 
\ee 
and recall the identity \eq{10}. \eq{17} is the second 
number zero in \eq{2}. But note that \eq{16} also shows 
what is left at order $g^4$.

All $g^4$ contributions in 1--loop diagrams are now collected 
from \eq{16}, \eq{11} and \eq{12}, the latter taken at 
$Q_0=P_0=0$. Even for the order $g^4$ of interest, the 
large--mass propagator $\Dm_\s$ in \eq{16} may be replaced 
by $\D_\s$. With $\,\Re\,$ denoting regulator terms, the 
result may be written as
\bea{18}
 \P_t^{\rm 1-loop} &=& g^2 N T \int^3_p \; \Bigg\{ \; 
      \; \Re \; + 
     {4\03}\,\D_\tau   + {1\02} \lammi \Dm_0 \D_0 
      \nonu \\ 
  & &  -\; {1\02} \lammi \( 3 + p^2 \Dm_0 \) 
      \Dm_\tau \D_\tau 
    - 3 q^2\lammi \D_0 \Dm_\tau \D_\tau  \nonu \\
  & &  - \;{1\04} q^4 \lammi \Dm_0 \D_0 \Dm_\tau \D_\tau
   \;\; + \quad \hbox{terms $\sim \a, \a^2$} 
   \;\;\; \Bigg\}  \quad . \quad 
\eea 
The regulators are
\be{19}
  \Re \; = \, \D_\s - 2 \D_M + {4\03}p^2 \D_M^2 
   - {2\03}p^2\D_\s^2 \, = \, 
   - \D_\s + {2\03}p^2\D_\s^2 \quad 
\ee 
with the right end valid for the choice $M=\s$. Hence, as the 
reminder in the wavy bracket \eq{18} behaves as ${1\03}\D_0$ 
for large $p$, the regulators prevent \eq{18} from diverging 
linearly. Remember, that independence on gauge--fixing can not 
be required, because now (being at order $g^4$) consistency
is only achieved by including the whole Linde sea. Therefore
the terms $\sim \a,\,\a^2$ in \eq{18}, which are UV finite,
 need not be detailed.

The expression \eq{18} is precisely what one obtains at 1--loop 
order for the 3D Yang--Mills theory at zero temperature and with 
coupling $e^2=g^2T$. We have done this 3D calculation in the 
metrics $+--$, by using the propagator $\MA^{\mu\nu} / (K^2-\tau^2) 
+ \a \MD^{\mu\nu} /K^2$, by rotating the zeroth momentum at the end 
and by tracing with ${1\02} \MA(Q)$. Because \eq{18} has order 
$g^4$ in magnitude, the 3D self--energy starts with $e^4\,$.

$ $   

{\bf 4. Two loops \ --- \ both inner momenta hard}

We turn to the third number zero in \eq{2}. Due to
its generic prefactor $g^4$, 2--loop diagrams with
hard inner momenta appear to give the natural contributions 
to the order $g^4$ of interest. In an other context, the 13 
diagrams of Figure 2 were analysed in \cite{nt}, but in 
the ''wrong'' limit ($q\to0$ first) and in Feynman gauge 
$\a=1$ only. Here, instead, we ask two other questions$\,$: 
first, whether the hard--hard diagrams form a gauge 
invariant subset in the static limit, and second, whether 
its contribution vanishes. Hence, lines are associated with 
the propagators $\,G_0(P) = g^{\mu\nu}/P^2 + (\a-1)P^\mu P^\nu/P^4$. 
Application of the diagrammatic rules and the colour sums 
were done by hand. But MAPLE programs were helpful in 
Lorentz contracting and sorting the various terms.

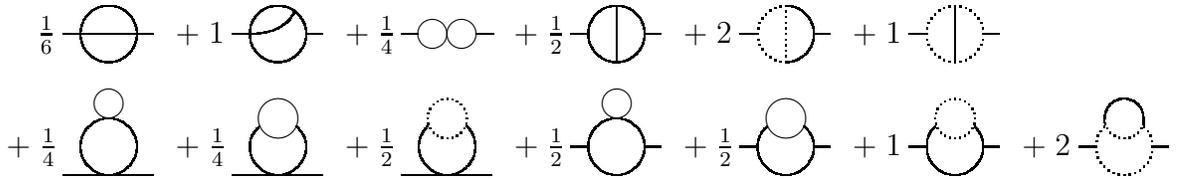
\begin{figure}[bth]  
\unitlength0.75cm \begin{picture}(19,3.8)
\put(0.55,2.35){${1\06}$} \put(1.0,2.5){\line(1,0){1.6}}    
\qbezier(1.8,3.0)(2.00711,3.0)(2.15355,2.85355)
\qbezier(2.15355,2.85355)(2.3,2.70711)(2.3,2.5)
\qbezier(2.3,2.5)(2.3,2.29289)(2.15355,2.14645)
\qbezier(2.15355,2.14645)(2.00711,2.0)(1.8,2.0)
\qbezier(1.8,2.0)(1.59289,2.0)(1.44645,2.14645)
\qbezier(1.44645,2.14645)(1.3,2.29289)(1.3,2.5)
\qbezier(1.3,2.5)(1.3,2.70711)(1.44645,2.85355)
\qbezier(1.44645,2.85355)(1.59289,3.0)(1.8,3.0)
\put(3.0,2.35){$+\;1$} \put(4.0,2.5){\line(1,0){0.3}} 
\put(5.3,2.5){\line(1,0){0.3}}  
\qbezier(4.8,3.0)(5.00711,3.0)(5.15355,2.85355)
\qbezier(5.15355,2.85355)(5.3,2.70711)(5.3,2.5)
\qbezier(5.3,2.5)(5.3,2.29289)(5.15355,2.14645)
\qbezier(5.15355,2.14645)(5.00711,2.0)(4.8,2.0)
\qbezier(4.8,2.0)(4.59289,2.0)(4.44645,2.14645)
\qbezier(4.44645,2.14645)(4.3,2.29289)(4.3,2.5)
\qbezier(4.3,2.5)(4.3,2.70711)(4.44645,2.85355)
\qbezier(4.44645,2.85355)(4.59289,3.0)(4.8,3.0)
\qbezier(4.3,2.5)(4.9,2.5)(5.1,2.9)
\put(6.0,2.35){$+\; {1\04}$}  \put(7.0,2.5){\line(1,0){0.3}}   
\put(8.33,2.5){\line(1,0){0.3}} \put(7.55,2.5){\circle{0.5}}    
\put(8.07,2.5){\circle{0.5}}
\put(9.0,2.35){$+\; {1\02}$} \put(10.0,2.5){\line(1,0){0.3}}
\put(11.3,2.5){\line(1,0){0.3}} \put(10.815,2.0){\line(0,1){1}}
\qbezier(10.8,3.0)(11.00711,3.0)(11.15355,2.85355)
\qbezier(11.15355,2.85355)(11.3,2.70711)(11.3,2.5)
\qbezier(11.3,2.5)(11.3,2.29289)(11.15355,2.14645)
\qbezier(11.15355,2.14645)(11.00711,2.0)(10.8,2.0)
\qbezier(10.8,2.0)(10.59289,2.0)(10.44645,2.14645)
\qbezier(10.44645,2.14645)(10.3,2.29289)(10.3,2.5)
\qbezier(10.3,2.5)(10.3,2.70711)(10.44645,2.85355)
\qbezier(10.44645,2.85355)(10.59289,3.0)(10.8,3.0)
\put(12.0,2.35){$+\; 2$} \put(13.0,2.5){\line(1,0){0.3}}   
\put(14.3,2.5){\line(1,0){0.3}}  
\qbezier(13.8,3.0)(14.00711,3.0)(14.15355,2.85355)
\qbezier(14.15355,2.85355)(14.3,2.70711)(14.3,2.5)
\qbezier(14.3,2.5)(14.3,2.29289)(14.15355,2.14645)
\qbezier(14.15355,2.14645)(14.00711,2.0)(13.8,2.0)
\put(15.0,2.35){$+\; 1$} \put(16.0,2.5){\line(1,0){0.3}}   
\put(17.3,2.5){\line(1,0){0.3}} \put(16.815,2){\line(0,1){1}}
\put(0.0,.35){$+\; {1\04}$} \put(1.0,.0){\line(1,0){1.6}}   
\put(1.815,1.265){\circle{0.5}}   
\qbezier(1.8,1.0)(2.00711,1.0)(2.15355,.85355)
\qbezier(2.15355,.85355)(2.3,.70711)(2.3,.5)
\qbezier(2.3,.5)(2.3,.29289)(2.15355,.14645)
\qbezier(2.15355,.14645)(2.00711,.0)(1.8,.0)
\qbezier(1.8,.0)(1.59289,.0)(1.44645,.14645)
\qbezier(1.44645,.14645)(1.3,.29289)(1.3,.5)
\qbezier(1.3,.5)(1.3,.70711)(1.44645,.85355)
\qbezier(1.44645,.85355)(1.59289,1.0)(1.8,1.0)
\put(3.0,.35){$+\; {1\04}$} \put(4.0,.0){\line(1,0){1.6}}  
\put(4.815,1.0){\circle{0.7}}
\qbezier(5.13000,.85355)(5.3,.70711)(5.3,.5)
\qbezier(5.3,.5)(5.3,.29289)(5.15355,.14645)
\qbezier(5.15355,.14645)(5.00711,.0)(4.8,.0)
\qbezier(4.8,.0)(4.59289,.0)(4.44645,.14645)
\qbezier(4.44645,.14645)(4.3,.29289)(4.3,.5)
\qbezier(4.3,.5)(4.3,.70711)(4.47000,.85355)
\put(6.0,.35){$+\; {1\02}$} \put(7.0,.0){\line(1,0){1.6}}   
\qbezier(8.15355,.85355)(8.3,.70711)(8.3,.5)
\qbezier(8.3,.5)(8.3,.29289)(8.15355,.14645)
\qbezier(8.15355,.14645)(8.00711,.0)(7.8,.0)
\qbezier(7.8,.0)(7.59289,.0)(7.44645,.14645)
\qbezier(7.44645,.14645)(7.3,.29289)(7.3,.5)
\qbezier(7.3,.5)(7.3,.70711)(7.44645,.85355)
\put(9.0,.35){$+\; {1\02}$} \put(10.0,.5){\line(1,0){0.3}}   
\put(11.3,.5){\line(1,0){0.3}} \put(10.815,1.265){\circle{0.5}}   
\qbezier(10.8,1.0)(11.00711,1.0)(11.15355,.85355)
\qbezier(11.15355,.85355)(11.3,.70711)(11.3,.5)
\qbezier(11.3,.5)(11.3,.29289)(11.15355,.14645)
\qbezier(11.15355,.14645)(11.00711,.0)(10.8,.0)
\qbezier(10.8,.0)(10.59289,.0)(10.44645,.14645)
\qbezier(10.44645,.14645)(10.3,.29289)(10.3,.5)
\qbezier(10.3,.5)(10.3,.70711)(10.44645,.85355)
\qbezier(10.44645,.85355)(10.59289,1.0)(10.8,1.0)
\put(12.0,.35){$+\; {1\02}$} \put(13.0,.5){\line(1,0){0.3}}  
\put(14.3,.5){\line(1,0){0.3}} \put(13.82,1.0){\circle{0.7}}
\qbezier(14.13000,.85355)(14.3,.70711)(14.3,.5)
\qbezier(14.3,.5)(14.3,.29289)(14.15355,.14645)
\qbezier(14.15355,.14645)(14.00711,.0)(13.8,.0)
\qbezier(13.8,.0)(13.59289,.0)(13.44645,.14645)
\qbezier(13.44645,.14645)(13.3,.29289)(13.3,.5)
\qbezier(13.3,.5)(13.3,.70711)(13.47000,.85355)
\put(15.0,.35){$+\; 1$} \put(16.0,.5){\line(1,0){0.3}}   
\put(17.3,.5){\line(1,0){0.3}}
\qbezier(17.15355,.85355)(17.3,.70711)(17.3,.5)
\qbezier(17.3,.5)(17.3,.29289)(17.15355,.14645)
\qbezier(17.15355,.14645)(17.00711,.0)(16.8,.0)
\qbezier(16.8,.0)(16.59289,.0)(16.44645,.14645)
\qbezier(16.44645,.14645)(16.3,.29289)(16.3,.5)
\qbezier(16.3,.5)(16.3,.70711)(16.44645,.85355)
\put(18.0,.35){$+\; 2$} \put(19.0,.5){\line(1,0){0.3}}   
\put(20.3,.5){\line(1,0){0.3}}  
\qbezier(19.80,1.35)(19.94497,1.35)(20.04749,1.24749)
\qbezier(20.04749,1.24749)(20.15,1.14497)(20.15,.93)
\qbezier(19.45,.93)(19.45,1.14487)(19.55251,1.24749)
\qbezier(19.55251,1.24749)(19.65503,1.35)(19.80,1.35)
\linethickness{.6pt}
\qbezier[10](13.8,2.0)(13.8,2.5)(13.8,3.0)
\qbezier[4](13.8,2.0)(13.59289,2.0)(13.44645,2.14645)
\qbezier[4](13.44645,2.14645)(13.3,2.29289)(13.3,2.5)
\qbezier[4](13.3,2.5)(13.3,2.70711)(13.44645,2.85355)
\qbezier[4](13.44645,2.85355)(13.59289,3.0)(13.8,3.0)
\qbezier[4](16.8,3.0)(17.00711,3.0)(17.15355,2.85355)
\qbezier[4](17.15355,2.85355)(17.3,2.70711)(17.3,2.5)
\qbezier[4](17.3,2.5)(17.3,2.29289)(17.15355,2.14645) 
\qbezier[4](17.15355,2.14645)(17.00711,2.0)(16.8,2.0)
\qbezier[4](16.8,2.0)(16.59289,2.0)(16.44645,2.14645)
\qbezier[4](16.44645,2.14645)(16.3,2.29289)(16.3,2.5)
\qbezier[4](16.3,2.5)(16.3,2.70711)(16.44645,2.85355)
\qbezier[4](16.44645,2.85355)(16.59289,3.0)(16.8,3.0)
\qbezier[3](7.80,1.35)(7.94497,1.35)(8.04749,1.24749)
\qbezier[3](8.04749,1.24749)(8.15,1.14497)(8.15,1.00)
\qbezier[3](8.15,1.00)(8.15,.85503)(8.04749,.75251)
\qbezier[3](8.04749,.75251)(7.94497,.65)(7.80,.65)
\qbezier[3](7.80,.65)(7.65503,.65)(7.55251,.75251)
\qbezier[3](7.55251,.75251)(7.45,.85503)(7.45,1.00)
\qbezier[3](7.45,1.00)(7.45,1.14487)(7.55251,1.24749)
\qbezier[3](7.55251,1.24749)(7.65503,1.35)(7.80,1.35)
\qbezier[3](16.80,1.35)(16.94497,1.35)(17.04749,1.24749)
\qbezier[3](17.04749,1.24749)(17.15,1.14497)(17.15,1.00)
\qbezier[3](17.15,1.00)(17.15,.85503)(17.04749,.75251)
\qbezier[3](17.04749,.75251)(16.94497,.65)(16.80,.65)
\qbezier[3](16.80,.65)(16.65503,.65)(16.55251,.75251)
\qbezier[3](16.55251,.75251)(16.45,.85503)(16.45,1.00)
\qbezier[3](16.45,1.00)(16.45,1.14487)(16.55251,1.24749)
\qbezier[3](16.55251,1.24749)(16.65503,1.35)(16.80,1.35)
\qbezier[4](20.15355,.85355)(20.3,.70711)(20.3,.5)
\qbezier[4](20.3,.5)(20.3,.29289)(20.15355,.14645)
\qbezier[4](20.15355,.14645)(20.00711,.0)(19.8,.0)
\qbezier[4](19.8,.0)(19.59289,.0)(19.44645,.14645)
\qbezier[4](19.44645,.14645)(19.3,.29289)(19.3,.5)
\qbezier[4](19.3,.5)(19.3,.70711)(19.44645,.85355)
\qbezier[3](20.15,1.00)(20.15,.85503)(20.04749,.75251)
\qbezier[3](20.04749,.75251)(19.94497,.65)(19.80,.65)
\qbezier[3](19.80,.65)(19.65503,.65)(19.55251,.75251)
\qbezier[3](19.55251,.75251)(19.45,.85503)(19.45,1.00)
\end{picture}
\caption{\ft The two--loop diagrams with symmetry factors 
in front of each. Dotted lines refer to ghost propagators, 
normal lines to gluons. }
\vskip .3cm\end{figure}

For convenience the result was split according to
powers of $(\a -1)$ and traced with $\MA$. Terms
$\sim (\a -1)^5$ occur in only two diagrams and drop out 
in each, immediately. Terms
$\sim (\a-1)^{4 \; {\rm to}\; 2}$ were seen to vanish on
the transversal line in the static limit, i.e. at $Q_0=0$
 and $q\to0$ in numerators. This somewhat laborious step 
involves several symmetry arguments and, finally, the 
suppression of all $q^2$ or $\vc q$, which remained
in numerators. The powers $(\a-1)^1$ and $(\a-1)^0$ 
require more detail. By working as indicated above we first 
obtain
\be{20}
  \P_t^{\rm 2-loop \; hh} \Big|_{\sim (\a-1)} =
  2\, g^4 N^2 \,\Big( - J_0(0)\, I_0 - J_0(0)\, I_1(Q) 
  - I_0\, J_1(Q) + 2\, L(Q)\,\Big) \quad 
\ee 
with 
\be{21}
  I_0 =\sum_P {1\0P^2} = -{T^2\012} \;\; , \;\quad
  J_0(Q)=\sum_P {1\0P^2(P-Q)^2} \quad 
\ee 
and
\bea{22}
   I_1(Q) &=&\sum_P {\lamm\0P^2(P-Q)^2} \;\; ,\;\quad
   J_1(Q)=\sum_P {\lamm\0P^4(P-Q)^2} \;\; , \nonu \\
  L(Q) &=& \sum_P\sum_K {(PK) \,\big[\, \vc p \vc k 
    - {(\vcsm p \vcsm q)(\vcsm k \vcsm q)\0q^2} \,\big]\, \0
        P^2 (P-Q)^2 K^4 (K-Q)^2 } \; = \;
      - {1\02} I_1 (Q) J_1(Q) \quad , \quad
\eea 
where a bit of spherical trigonometry has led to the last 
expression. The remaining angular integrations can be done
as well. Now, in a last step, the limit $\o\to0$ (first) and
$q \to 0$ (afterwards) is performed. The subleties of
this limit are detailed in the Appendix, where we obtain 
that $\,I_1(Q) \to - I_0\,$. Hence 
\be{23}
  \P_t^{\rm 2-loop \; hh} \Big|_{\sim (\a-1)} =
  -2\, g^4 N^2 \,\Big(\, J_0(0) + J_1(Q) \,\Big) \,\Big(\,
   I_0 + I_1(Q)\,\Big) \, \to \, 0 \;\; . \quad 
\ee 
Thus, in this limit, the hard--hard 2--loop contribution 
is a gauge--fixing independent subset, indeed.

The contribution itself, which is now given by the $(\a-1)^0$ 
term, is obtained to be
\be{24}
  \P_t^{\rm 2-loop \; hh} =  - 4\, g^4 N^2 \,\Big(\, I_1(Q)\, 
  J_0(Q) -2 I_0\, J_1(Q) \,\Big) 
  \; \to \; 0 \;\;
\ee 
by virtue of $-2J_1(Q) \to J_0(Q)$ in the static limit.
\eq{24} is the third number zero in \eq{2}.

$ $   

{\bf 5. Two loops \ --- \ one hard, one supersoft}

There is, of course, also the $g^4$ contribution in the 2--loop 
diagrams, which arises from the supersoft region (of both 
momenta) and may be addressed to the 3D Euklidean theory. Due 
to the vanishing of the hard--hard part, we may assume, that 
the Euclidean part is anyhow regularized. The very details 
of this behaviour are under present study.

One may suspect that a contribution is omitted so far, namely 
the admixture of one loop momentum hard and the other supersoft 
and regularized. The system might ''know'' of a soft scale only 
through these regulators. Such contributions
$^2\hspace*{-.8mm}\P_t^{\rm hard-supersoft}$, however, can be 
ruled out either (a) by taking into account 
the one--loop--higher subtraction of $\ov{Y}$ terms or
(b) by power counting. 

For the mechanism (a), consider the first six diagrams in the 
second line of figure 2 with the lower loop at soft momentum. 
They combine with 1--loop diagrams containing a $\ov{Y}$ 
insertion. The sum of these diagrams vanishes to order $g^4$.

For the power counting (b), consider for example the setting 
sun diagram, which is the first in figure 2. Reducing one 
loop integration to the zero mode and supplying it with a 
Pauli--Villars regulator, one has 
\be{26}
  \sim \; g^4 \sum_K^{\rm hard} T \int^3_p {1\0 K^4} 
  \( {1\0p^2+\tau^2} - {1\0p^2+\s^2}\) \;\sim\; g^4 T \s 
  \;\sim\; g^5 T^2 \;\; ,  \quad
\ee 
hence one $g$ order below the one of present interest. 

It is now tempting to speculate on how to go to higher 
orders in the perturbative treatment. ''Towers'' (i.e.
HTL vertex insertions) in the Linde sea are to be included 
\ -- \ one in each diagram, if the order $g^5$ is studied. 
Their combinatorics of positions and the question for an 
effective action could be future problems.

$ $   
 
{\bf 6. Conclusions}

The magnetic screening mass has no other contributions
than those with zero Matsubara frequency in each loop. 
It is thus given, to its leading order, by the 
Karabali--Nair value $g^2 N T /(2\pi)\,$. The Linde problem 
is going to be overcome.

$ $   

\let\dq=\thq \renewcommand{\theequation}{A.\dq}    
\setcounter{equation}{0}          

{\bf Appendix}

Here the non--commutativity of limits, long--wavelength
versus static, is illustrated with the sum--integral 
\bea{a1}
 I_1(Q) &=& \sum_P { \lamm \0 P^2 (P-Q)^2 } \, = \,
 {- Q^2\0 4\pi^2 q^2} \int_0^\infty\!\! dp \; p\, n(p)\;
 \Bigg\{\; 1 - {Q_0\0q}\;\ln \( {Q_0+q \0 Q_0-q}\) - \nonu \\
 & &\hspace*{-1cm} - \; {Q_0\02q}\;
    \ln \( {(Q_0-q)^2 - 4p^2 \0 (Q_0+q)^2 -4p^2}\)\, 
    + \, {p\0 2q} \( 1 + {Q^2 \0 4p^2}\)\;
    \ln \( {Q_0^2 -(2p+q)^2 \0 Q_0^2 - (2p-q)^2} \)
    \; \Bigg\} \; . \qquad
\eea 
at hand. A temperature independent piece was omitted to the 
right and addressed to renormalization at $T=0$. $Q_0$ may 
still attain the Matsubara values. Analytical 
continuation $Q_0 = \o + i \eta$ reveals several cuts 
on the real axis. Maintaining $\pm Q_0$ symmetry, one cut 
inevitably extends from $\o=-q$ to $\o=q$ (if $q$ is real). 
The others lie outside, with inner end points of order
$p \sim T$ in magnitude. Hence, when concentrating on soft 
or supersoft $\o$, $q$, \eq{a1} may be simplified to
\be{a2}
  I_1(\o+i\eta, \vc q) = {T^2\024} \lk\; 2 
  - {\,\o+i\eta \,\0 q} \,\ln \( {\o + q + i\eta 
  \0 \o - q + i\eta } \) \;\rk \( 1 
  - { (\o+i\eta)^2\0 q^2} \)  \quad
\ee 
with the only cut from $-q$ to $q$. When reaching the
plasmon frequency $\o \to m$ with $q \to 0$ ($\eta=0$),
\eq{a2} attains the ''usual'' value of $T^2/36$.
For negative $q^2$, however, the cut extends along the 
imaginary axis. The limit $\o \to 0$ (again $\eta=0$) is 
now possible, while $|q|$ remains finite, though supersoft. 
Hence, in the static limit, \eq{a2} becomes $T^2/12$.
This is also the value of $I_1$ at $Q_0=0$ ($|q| \ll T$) 
before continuing analytically.

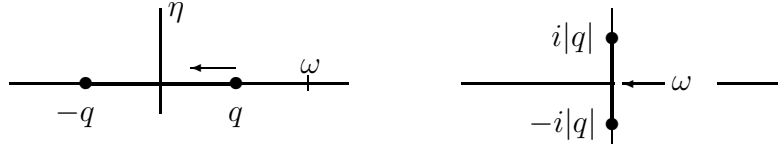
\begin{figure}[bth]  \unitlength 1cm  
\begin{picture}(11,1.4)
\put(2,0){\line(1,0){4.5}} \put(4,-.4){\line(0,1){1.4}}  
\put(2.9,-.1){$\bullet$}   \put(4.9,-.1){$\bullet$}
\put(4.1,.9){$\eta$} \put(5.86,.14){$\o$}
\put(5.96,-.1){\line(0,1){.2}}
\put(4.9,-.5){$q$}  \put(2.6,-.5){$-q$}
\put(5,.2){\vector(-1,0){.6}}
\thicklines  \put(3,-.01){\line(1,0){2}} \thinlines
\put(8,0){\line(1,0){2.04}}  \put(11.4,0){\line(1,0){.8}}   
\put(10,-.8){\line(0,1){1.8}}  
\put(9.9,-.64){$\bullet$} \put(9.9,.5){$\bullet$}
\put(8.9,-.64){$-i|q|$} \put(9.2,.5){$i|q|$}
\put(10.7,0){\vector(-1,0){.55}} \put(10.8,-.08){$\o$}
\thicklines  \put(10,-.6){\line(0,1){1.2}}  \thinlines
\end{picture} \vspace*{.5cm}
\caption{\ft The complex $Q_0 = \o+i\eta$ plane and 
the two limits encountered at long wavelength with
the real dispersion (left part of the figure) and in 
the static limit (right part).}
\end{figure}


\renewcommand{\section}{\paragraph}

\end{document}